%% file: paper_corr_recon_arxiv.tex
\documentclass[review,12pt]{elsarticle}
\usepackage[font=footnotesize,caption=false]{subfig}
\usepackage{endfloat}
\usepackage[utf8]{inputenc}
\usepackage{ifthen}
\makeatletter
\@ifpackageloaded{hyperref}{}{\usepackage[colorlinks,linkcolor=black,citecolor=black,urlcolor=black]{hyperref}}
\makeatother
\usepackage[round-mode=figures,binary-units]{siunitx}
\DeclareSIUnit\sample{sample}
\DeclareSIUnit[per-mode=symbol]\bpsamp{\bit\per\sample}
\usepackage{xfrac}
\usepackage{booktabs}
\usepackage[acronym]{glossaries}
\loadglsentries{acronyms.tex}
\glsdisablehyper

\usepackage{amsmath,amssymb,amsthm,xspace,bm}
\newtheorem{prop}{Proposition}
\makeatletter
\let\qnoise\@undefined
\makeatother
\input{cs_math.tex}
\newcommand{\csuncorqnsymb}{w}
\newcommand{\csuncorqn}[1][]{\ensuremath{\mathbf \csuncorqnsymb_{#1}}\xspace}
\newcommand{\bpdnsolset}[1]{\ensuremath{\uppercase{#1}}\xspace}
\newcommand{\argmin}{\operatornamewithlimits{argmin}}
\newcommand{\Argmin}{\operatornamewithlimits{Argmin}}

\journal{EURASIP Signal Processing}
\begin{document}
\begin{frontmatter}
  

  
  \title{Compressed Sensing with Linear Correlation Between Signal and
    Measurement Noise}

  \author{Thomas Arildsen\textsuperscript{a,$\ast$} and Torben
    Larsen\textsuperscript{b}}

  \address{Aalborg University, Faculty of Engineering and Science\\
    Department of Electronic Systems\\
    Postal address:\\
    Niels Jernes Vej 12, DK-9220 Aalborg, Denmark\\
    ---\\
    \begin{tabular}{ll}
      \textsuperscript{a}e-mail: tha@es.aau.dk & \textsuperscript{b}e-mail: tl@es.aau.dk\\
      phone: +45 99409844\\
      ORCID: 0000-0003-3254-3790\\
      \textsuperscript{$\ast$}(Corresponding author, EURASIP member)
    \end{tabular}
  }

\begin{abstract}
  Existing convex relaxation-based approaches to reconstruction in
  compressed sensing assume that noise in the measurements is
  independent of the signal of interest. We consider the case of noise
  being linearly correlated with the signal and introduce
  a simple technique for improving compressed sensing
  reconstruction from such measurements.  The technique is based on a
  linear model of the correlation of additive noise with the
  signal. The modification of the reconstruction algorithm based on
  this model is very simple and has negligible additional
  computational cost compared to standard reconstruction algorithms,
  but is not known in existing literature. The proposed
  technique reduces reconstruction error considerably in the case of
  linearly correlated measurements and noise. Numerical experiments
  confirm the efficacy of the technique. The technique is demonstrated
  with application to low-rate quantization of compressed
  measurements, which is known to introduce correlated noise, and
  improvements in reconstruction error compared to ordinary
  \acrlong{BPDN} of up to approximately 7\:dB are observed for
  1\:bit/sample quantization. Furthermore, the proposed method is
  compared to \acrlong{BIHT} which it is demonstrated to outperform in
  terms of reconstruction error for sparse signals with a number of
  non-zero coefficients greater than approximately \sfrac{1}{10}th of
  the number of compressed measurements.
\end{abstract}


\end{frontmatter}

\section{Introduction}
\label{sec:introduction}

In the recently emerged field of compressed sensing,
one considers linear measurements \csmeas of a sparse vector
\cssparsevec, possibly affected by noise as:
\begin{equation}
  \label{eq:basic-cs-meas}
  \csmeas = \cssysmtx\cssparsevec + \noise,
\end{equation}
where the measurements $\csmeas \in \mathbb R^{M\times 1}$, the sparse
vector $\cssparsevec \in \mathbb R^{N\times 1}$, the additive noise
$\noise \in \mathbb R^{M\times 1}$, the system matrix $\cssysmtx \in
\mathbb R^{M\times N}$, and $M \ll N$
\cite{Donoho2006,Candes2006,Candes2006b}.  \cssysmtx is generally the
product of a measurement matrix and a dictionary matrix: $\cssysmtx =
\csmeasmtx\csdict$, where $\csmeasmtx \in \mathbb C^{M\times N}$,
$\csdict \in \mathbb C^{N\times N}$. For simplicity, we assume that
\csdict is an orthonormal basis although more general dictionaries are
indeed possible~\cite{Candes2011}.

The essence of compressed sensing, as Donoho, Candès, Romberg, and Tao
show in \cite{Donoho2006,Candes2006}, is that the under-determined
equation system (\ref {eq:basic-cs-meas}) can be solved provided that:
\begin{enumerate}
\item The vector \cssparsevec is sparse; i.e., only few ($K$) elements
  in \cssparsevec are non-zero.
  \begin{equation}
    \label{eq:sparsity}
    K = \left|\left\{\cssparsevecsymb_i \middle| \cssparsevecsymb_i\neq
        0, i = 1,\ldots,N\right\}\right|
  \end{equation}
  \cssparsevec can also be approximated sparsely if it is compressible
  \cite[Sec. 3.3]{Candes2006b}, meaning that its coefficients sorted
  by magnitude decay rapidly to zero.
\item The system matrix \cssysmtx obeys the \gls{RIP} with isometry
  constant $\delta_K > 0$, defined as follows:
  \begin{equation}
    \label{eq:rip}
    \left(1-\delta_K\right)\left\|\cssparsevec\right\|_{\ell_2}^2 \leq
    \left\|\cssysmtx\cssparsevec\right\|_{\ell_2}^2 \leq
    \left(1+\delta_K\right)\left\|\cssparsevec\right\|_{\ell_2}^2,
  \end{equation}
  for any at most $K$-sparse vector
  \cssparsevec such that~\cite{Candes2005}:
  \begin{equation}
    \label{eq:rip-delta-bound}
    \delta_K + \delta_{2K} + \delta_{3K} < 1.
  \end{equation}
  This holds with high probability when \csmeasmtx is generated with
  zero-mean \gls{IID} Gaussian entries with variance $\frac{1}{M}$.
  Note that (\ref {eq:rip}) and~(\ref {eq:rip-delta-bound}) are sufficient but not
  necessary conditions, and rather conservative conditions indeed, as
  shown in~\cite{Donoho2010}.
  
  Conditions~(\ref {eq:rip}) and~(\ref {eq:rip-delta-bound}) lead to the following
  sufficient amount of measurements $M$ for Gaussian measurement
  matrices \csmeasmtx~\cite{Candes2006a}:
  \begin{equation}
    \label{eq:enough-meas-rip}
    M \geq C K \log\left(\frac{N}{M}\right),
  \end{equation}
  where $C$ is a fairly small constant which can be calculated as a
  function of
  $\frac{M}{N}$~\cite{Candes2005}.
\end{enumerate}

Given the measurements \csmeas, the unknown sparse vector \cssparsevec
can be reconstructed by solving the following convex optimization
problem~\cite[Sec. 4]{Candes2006b}:
\begin{equation}
  \label{eq:cvx-sparse-sol}
  \cssparsevecest  = \argmin_{\cssparsevecvar:\
    \|\csmeas-\cssysmtx\cssparsevecvar\|_2 \le \epsilon}
  \|\cssparsevecvar\|_1,
\end{equation}
where the fidelity constraint $\|\csmeas-\cssysmtx\cssparsevecvar\|_2
\le \epsilon$ ensures consistency with the observed measurements to
within some margin of error, $\epsilon$, which is chosen sufficiently
large to accommodate the error $\noise$ and/or approximation error
in the case of compressible signals.
% We refer to Section~\ref {sec:proposed-approach} for further details on the
% choice of $\epsilon$.
The form of the optimization problem in (\ref {eq:cvx-sparse-sol}) is
known as \gls{LASSO} \cite{Tibshirani1996} or \gls{BPDN}
\cite{Chen1998} and also comes in other variants such as the Dantzig
selector~\cite{Candes2007a}. In addition to the convex optimization
approach to reconstruction in compressed sensing, there exist several
iterative/greedy algorithms such as \gls{IHT}~\cite{Blumensath2010a},
or \gls{SP}~\cite{Dai2009b} and \gls{COSAMP}~\cite{Needell2009a} as
well as the more generalized incarnation of the two latter,
\gls{TST}~\cite{Maleki2010}. We generally refer to such convex or
greedy approaches as reconstruction algorithms. The reconstruction
algorithms generally assume the noise to be white and independent of
the measurements before noise $\bar\csmeas = \cssysmtx\cssparsevec$. In
particular, to the best of the authors' knowledge, the case of
measurement noise being linearly correlated with the measurements has not been
treated in the existing literature. Such correlation arises in for
example the case of low-resolution quantization. As we demonstrate in
Section~\ref {sec:methodology}, this case poses a problem for the accuracy of
the found solution \cssparsevecest. More special cases of correlated
noise arising from Poisson measurements or quantisation of
measurements has, however, been treated in for example
\cite{Raginsky2010,Jacques2011,Jacques2013}.

In this paper, we propose a simple yet efficient approach to
alleviating the problem of linear correlation between the
measurements before noise $\bar\csmeas$ and the noise \noise. Our proposal boils
down to a simple scaling of the solution \cssparsevecest. Through
numerical experiments we demonstrate how linearly correlated measurements and
noise adversely affect the reconstruction error and demonstrate how
our proposal improves the estimates considerably.

As an application example, we demonstrate the proposed approach in the
case of low-rate scalar quantization of the measurements
$\bar\csmeas$ which can be observed to introduce the
mentioned linearly correlated measurement noise. We demonstrate how a
well-known linear model used for modeling such correlation in scalar
quantization is equivalent to the model of correlated measurement
noise considered in this work.

The article is structured as follows: Section~\ref {sec:methodology}
introduces the considered model of linear correlation between compressed
measurements and noise and proposes a solution to enhance
reconstruction under these conditions, Section~\ref {sec:simulation-results}
describes simulations conducted to evaluate the performance of the
proposed approach compared to a traditional approach,
Section~\ref {sec:numer-sim-results} presents the results of these numerical
simulations, Section~\ref {sec:discussion} provides discussions of some of the
presented results, and Section~\ref {sec:conclusion} concludes the article.

\section{Methodology}
\label{sec:methodology}

\subsection{Correlated Measurements and Noise}
\label{sec:corr-meas-noise}
We consider additive measurement noise \noise which is correlated with
the measurements before noise $\bar\csmeas$. We model the correlation
by the linear model:
\begin{equation}
  \label{eq:correlation-linear}
  \csmeas = \alpha\cssysmtx\cssparsevec + \csuncorqn,
\end{equation}
where \csuncorqn is assumed an additive white noise uncorrelated with
\cssparsevec and $0 < \alpha \le 1$ where $\alpha = 1$ covers the
ordinary case of uncorrelated measurement noise. \cssysmtx is the
product of a measurement matrix \csmeasmtx with \gls{IID} Gaussian
entries $\sim\mathcal{N}\left(0,\frac{1}{M}\right)$ and an orthonormal
dictionary matrix \csdict. The model (\ref {eq:correlation-linear})
results in the following additive noise term:
\begin{equation}
  \noise = \csmeas - \bar\csmeas
  = \alpha\cssysmtx\cssparsevec + \csuncorqn - \cssysmtx\cssparsevec
  = (\alpha-1) \cssysmtx\cssparsevec + \csuncorqn \label{eq:correlated-noise}
\end{equation}
We define $\bar\csmeas = \cssysmtx\cssparsevec$ to signify the
measurements before introduction of additive noise.  It is readily
seen from (\ref {eq:correlated-noise}) that \noise is correlated with
\cssparsevec.
The noise variance is
\begin{equation}
  \label{eq:correlated-noise-var}
  \sigma_\noisesymb^2 = \frac{1}{M} \expectation{\noise\transpose\noise}
  = \frac{1}{M}\left((\alpha - 1)^2 \expectation{\bar\csmeas\transpose
      \bar\csmeas} + \expectation{\csuncorqn\transpose
      \csuncorqn}\right),
\end{equation}
which can be calculated by assuming that $\sigma_{\bar\csmeassymb}^2 =
\frac{1}{M} \expectation{\bar\csmeas\transpose \bar\csmeas}$ and
$\sigma_{\csuncorqnsymb}^2 = \frac{1}{M}
\expectation{\csuncorqn\transpose \csuncorqn}$ are known or can be
estimated. For example, %we show in \cref{sec:pract-estim-noise} how
%$\sigma_{\bar\csmeassymb}^2$ can be estimated. Likewise,
we show an
example for $\sigma_{\csuncorqnsymb}^2$ in the case of quantization in
Section~\ref {sec:an-appl-quant}, (\ref {eq:r-noise-var}).

The specific problem caused by correlated measurements and noise as
modeled by (\ref {eq:correlation-linear}) is that the noise itself is
partly sparse in the same dictionary as the signal of interest,
\cssparsevec. Intuitively, this causes a solution \cssparsevecest as
given by, e.g., (\ref {eq:cvx-sparse-sol}) to adapt to part of the noise
\emph{as well as} the signal of interest, unless steps are taken to
mitigate this effect.

\subsection{Proposed Approach}
\label{sec:proposed-approach}

Using the model in (\ref {eq:correlation-linear}), we propose the
following reconstruction of the sparse vector \cssparsevec instead of
the standard approach in
(\ref {eq:cvx-sparse-sol}). Equation~(\ref {eq:correlation-linear}) motivates
replacing the system matrix \cssysmtx by its scaled version
$\alpha\cssysmtx$. We exemplify this approach by applying it in the
\gls{BPDN} reconstruction formulation as below. Replacing \cssysmtx by
$\alpha\cssysmtx$ in the standard approach (\ref {eq:cvx-sparse-sol}),
we arrive at
\begin{equation}
  \label{eq:bpdn-proposed}
  \cssparsevecest[1] = \argmin_{\cssparsevecvar:\
    \|\csmeas-\alpha\cssysmtx\cssparsevecvar\|_2 \le
    \epsilon} \|\cssparsevecvar\|_1.
\end{equation}

Since $\epsilon$ should be chosen to accommodate the level of noise in
the measurements \csmeas, we can see that, one choice could be to set
\begin{align}
  \epsilon &= \|\noise\|_2 \label{eq:straightforward-epsilon-n}\\
  \epsilon &= \|\csuncorqn\|_2\label{eq:straightforward-epsilon-r}
\end{align}
in (\ref {eq:cvx-sparse-sol}) or (\ref {eq:bpdn-proposed}),
respectively. Since the noise terms \noise and \csuncorqn are assumed
unknown,
(\ref {eq:straightforward-epsilon-n}) and~(\ref {eq:straightforward-epsilon-r}) are
not realistic choices of $\epsilon$. The optimal choice of $\epsilon$
is dependent on the true solution \cssparsevec, and is therefore
difficult to obtain in practice as exemplified for more general
inverse problems in, e.g., \cite{Hansen1987}. For this reason, various
rules of thumb exist for the selection of $\epsilon$. One such choice
is found in~\cite[Sec. 5.3]{Becker2011}:
\begin{equation}
  \label{eq:epsilon-choice-becker}
  \epsilon = \sqrt{M+2\sqrt{2M}}\sigma,
\end{equation}
where $\sigma$ is the noise level (standard deviation) of the
stochastic error \noise or \csuncorqn in (\ref {eq:basic-cs-meas}) or
(\ref {eq:correlation-linear}), respectively.

\subsection{Additional Insight on the Proposed Approach}
\label{sec:addit-insight-prop}

As outlined in Section~\ref {sec:proposed-approach}, the model of the
correlation between \noise and $\bar\csmeas$ suggests scaling
\cssysmtx in the constraint of (\ref {eq:bpdn-proposed}). In fact, as we
show here, an equivalent solution can be obtained simply by scaling the
solution found by the optimization formulation
(\ref {eq:cvx-sparse-sol}).
\begin{prop}
  \label{prop:equivalent-solutions}
  The following optimization formulation is equivalent to the
  formulation (\ref {eq:bpdn-proposed}) in the sense that they produce
  solutions of comparable precision:
  \begin{equation}
    \label{eq:bpdn-postscale}
    \cssparsevecest[2] = \frac{1}{\alpha} \argmin_{\cssparsevecvaralt:\
      \|\csmeas-\cssysmtx\cssparsevecvaralt\|_2 \le \epsilon}
    \|\cssparsevecvaralt\|_1.
  \end{equation}
\end{prop}
To see why (\ref {eq:bpdn-postscale}) is equivalent to
(\ref {eq:bpdn-proposed}), consider the optimization problem over the
variable \cssparsevecvaralt, in which we introduce a change of
variable $\cssparsevecvaralt \curvearrowright \cssparsevecvar$:
\begin{equation}
  \label{eq:bpdn-proposed1}
  \begin{aligned}
    \cssparsevecest[2] \in \bpdnsolset{\hat x_2} &=
    \Argmin_{\cssparsevecvaralt:\
      \|\csmeas-\cssysmtx\cssparsevecvaralt\|_2 \le
      \epsilon} \left\|\frac{1}{\alpha} \cssparsevecvaralt\right\|_1\\
    &= \Argmin_{\cssparsevecvar:\
      \|\csmeas-\alpha\cssysmtx\cssparsevecvar\|_2 \le \epsilon}
    \|\cssparsevecvar\|_1,\ \cssparsevecvaralt =
    \alpha\cssparsevecvar\\
    &= \bpdnsolset{\hat x_1} \ni \cssparsevecest[1]
  \end{aligned}
\end{equation}
In (\ref {eq:bpdn-proposed1}) we use the notation $\bpdnsolset{\hat x} =
\Argmin \ldots$ to denote the set of solutions to the stated
optimization problem since this is generally not one unique solution
\cite[Ch. 5]{Elad2010}. $\cssparsevecest \in \bpdnsolset{\hat x}$ is
used to emphasize that \cssparsevecest is any feasible minimizer of
the problem. It can generally not be guaranteed that algorithms used
to obtain solutions to the two optimization problems
(\ref {eq:bpdn-proposed}) and~(\ref {eq:bpdn-proposed1}) return the same solution,
but they are subject to the same guarantees of reconstruction accuracy
(stability) as given by \cite[Theorem 5.3]{Elad2010}.

According to the above, down-scaling the
solution to the optimization in (\ref {eq:bpdn-postscale}) by $\alpha$
results in a solution \cssparsevecest[2] of comparable accuracy to the solution
\cssparsevecest[1] to (\ref {eq:bpdn-proposed}).  Please note that all
constraints in
(\ref {eq:bpdn-proposed}), (\ref {eq:bpdn-postscale}) and~(\ref {eq:bpdn-proposed1}) use the
same value of $\epsilon$ given by (\ref {eq:epsilon-choice-becker}) with
$\sigma = \sigma_{\csuncorqnsymb}$, the standard deviation of the
entries in \csuncorqn{} in (\ref {eq:correlation-linear}).

In short, Proposition~\ref {prop:equivalent-solutions} says that for compressed
measurements with noise correlated with the measurements according to
the model (\ref {eq:correlation-linear}), given the correlation
parameter $\alpha$, when the signal \cssparsevec is reconstructed
using \gls{BPDN}, (\ref {eq:cvx-sparse-sol}), the obtained solution
should be scaled by the factor $\frac{1}{\alpha}$ to account for the
effect of the correlation.

\subsection{Optimality of the Proposed Approach}
\label{sec:optim-prop-appr}

In relation to the method proposed in
Sections~\ref {sec:proposed-approach} and~\ref {sec:addit-insight-prop}, it is of course
interesting to investigate whether the corrective scaling by $\alpha$
in the reconstruction of \cssparsevec is indeed optimal. To
investigate this, consider the following optimization formulation:
\begin{gather}
  \label{eq:bpdn-modified-beta}
  \cssparsevecest[\beta] = \frac{1}{\beta} \argmin_{\cssparsevecvar:\
    \|\csmeas-\cssysmtx\cssparsevecvar\|_2 \le \epsilon}
  \|\cssparsevecvar\|_1,
\end{gather}
where $\epsilon$ is given by (\ref {eq:epsilon-choice-becker}) and the
optimization problem is evaluated for a number of values of $\beta \in
[\alpha - \beta_1, \alpha + \beta_2]$ for a given value of $\alpha$
used in the correlated noise model (\ref {eq:correlation-linear}) and a
suitable choice of $\beta_1$ and $\beta_2$.  The numerical results of
this investigation can be found in
Section~\ref {sec:scale-and-reg-values}. $\beta = \alpha$ intuitively seems a
suitable choice, but numerical experiments indicate that it is in fact
not optimal. An explanation of this observation is offered in
Section~\ref {sec:discussion}.

\subsection{An Application: Quantization}
\label{sec:an-appl-quant}

As a practical example where the introduced measurement noise is
correlated with the measurements, we investigate low-rate scalar
quantization of the individual compressed measurements in \csmeas.
Quantization is usually modeled by an additive noise
model~\cite{Jayant1984}:
\begin{equation}
  \label{eq:additive-q}
  y = Q(\bar y) = \bar y + q,
\end{equation}
where $\bar y$ is the original value before quantization, which we
consider as $\bar y \in \mathbb R$. $Q(\cdot)$ is the (non-linear)
operation of scalar quantization, mapping $\bar y$ to an index $i$
representing a quantized value $y$
\begin{equation}
  \label{eq:quantization-operation}
  Q:\ \bar y \rightarrow y_i,\ \text{if}\ \bar y \in R_i, i \in
  \{1,\ldots,L\},%\\
  %&y = y_i \in \{y_1,y_2,\ldots,y_L\}\notag,
\end{equation}
where the range of input values is partitioned into $L$ regions $R_i,\
i \in \{1,\ldots,L\}$ and any value $\bar y \in R_i$ is quantized to
the point $y_i \in R_i$.  For input $\bar y$ with unbounded support,
the regions $R_i$ can be defined as follows:
\begin{equation}
  \label{eq:quantization-regions}
  R_i =
  \begin{cases}
    (p_{i-1},p_i],&\text{for}\ i=1,\ldots,L-1\\
    (p_{i-1},p_i),&\text{for}\ i=L,
  \end{cases}
\end{equation}
where $p_0 = -\infty \wedge p_L = \infty$. The additive noise $q=y -
\bar y$ represents the error introduced by quantizing $\bar y$ to the
value $y$.

Various modeling assumptions are typically made about $q$. One type of
quantizers has centroid codebooks, i.e. quantizers where the
reconstruction points $y_i$ are calculated as the respective centroids
of the distribution of the input $\csmeassymb$ in each of the regions
$R_i$, e.g., Lloyd-Max quantizers~\cite{Lloyd1982,Max1960}. For
quantizers with centroid codebooks, $q$ is correlated with the input
$x$.  A model of this correlation used in the literature is the
so-called gain-plus-additive-noise
model~\cite[Sec. II]{Westerink1992}:
\begin{equation}
  \label{eq:gain-plus-additive-q}
  y = Q(\bar y) = \alpha \bar y + r,
\end{equation}
where $\alpha \in [0,1]$ and $r$ is an additive noise, assumed
uncorrelated with $\bar y$. The variance of $\rnoisesymb$ is
\begin{equation}
  \label{eq:r-noise-var}
  \sigma^2_\rnoisesymb = \alpha(1-\alpha) \sigma^2_{\bar\csmeassymb}.
\end{equation}
The variance of $\qnoisesymb$ is
\begin{equation}
  \label{eq:q-noise-var}
  \sigma^2_\qnoisesymb = (1-\alpha) \sigma^2_{\bar\csmeassymb},
\end{equation}
which is easily seen by inserting (\ref {eq:r-noise-var}) in
$\sigma^2_\qnoisesymb = (\alpha-1)^2 \sigma_{\bar\csmeassymb} +
\sigma^2_\rnoisesymb$.

The parameter $\alpha$ can be computed for a specific quantizer. One
way to do this is to estimate it numerically by Monte-Carlo
simulation. From~\cite[Eq. (8)]{Westerink1992} we have
\begin{equation}
  \label{eq:alpha-estimate}
  \alpha = 1 - \frac{\sigma_q^2}{\sigma_{\bar\csmeassymb}^2}.
\end{equation}
The procedure is to generate a random test sequence $\bar\csmeassymb$,
quantize it with the given quantizer $Q$ designed\footnote{The
  quantizer can for example be trained on test data representing
  $\bar\csmeassymb$ or calculated based on the known or assumed
  \gls{PDF}\glsreset{PDF} of $\bar\csmeassymb$.} for the \gls{PDF} of
$\bar\csmeassymb$, estimate the variances $\sigma_q^2$ and
$\sigma_{\bar\csmeassymb}^2$ from the realizations of
$\bar\csmeassymb$ and $q=\bar\csmeassymb - y$, and use these to
calculate (\ref {eq:alpha-estimate}).

The model (\ref {eq:gain-plus-additive-q}) of the quantizer corresponds
to the proposed model of correlated measurements and noise described
by (\ref {eq:correlation-linear}), where $\rnoisesymb = \csuncorqnsymb$.
Please note that the model, (\ref {eq:gain-plus-additive-q}), considers
scalar quantization. In the case of quantization of a vector $\mathbf
v$, we use $Q(\mathbf v)$ to signify scalar quantization of the
individual elements of the vector $\mathbf v$.

We consider quantization of compressed measurements \csmeas of the
signal \cssparsevec:
\begin{align}
  \label{eq:cs-meas-quant}
  \csmeas &= Q\left(\cssysmtx \cssparsevec\right)\\
  \label{eq:cs-meas-quant-qcorrnoise}
  &= \cssysmtx\cssparsevec + \qnoise\\
  \label{eq:cs-meas-quant-new}
  &\approx \alpha\cssysmtx\cssparsevec + \csuncorqn,
  %\intertext{where}
  %&\cssparsevec = \csdict \cssparsevec,\notag
\end{align}
where
\begin{align*}
  %\label{eq:quant-noise-covs}
  \expectation{\qnoise\qnoise\transpose} &=
  \sigma^2_{\qnoisesymb}\mathbf I,&
  \expectation{\bar\csmeas\bar\csmeas\transpose} &=
  \sigma^2_{\bar\csmeassymb}\mathbf I,
\end{align*}
and $\mathbf I$ is the $M\times M$ identity matrix.

Approximating the quantization operation by the noise model in
(\ref {eq:cs-meas-quant-new}), we propose using the reconstruction
technique (\ref {eq:bpdn-postscale}) to improve reconstruction with
scalar quantized compressed measurements, (\ref {eq:cs-meas-quant}), as
an example of noise correlated with the measurements.

Noise variance estimates given by
(\ref {eq:r-noise-var}) and~(\ref {eq:q-noise-var}) can be obtained from a known
$\sigma^2_{\bar\csmeassymb}$. In hardware implementations,
$\sigma^2_{\bar\csmeassymb}$ can be considered known through the use
of automatic gain control prior to quantization or by other means of
estimating signal variance prior to quantization.

\section{Simulation Framework}
\label{sec:simulation-results}

\newlength{\figurewidth}
\newlength{\figureheight}

In this section we present the numerical simulation set-up used to
evaluate the reconstruction method proposed in
(\ref {eq:bpdn-postscale}).

\begin{figure}[!t]
  \setlength{\figurewidth}{.85\columnwidth}
  \setlength{\figureheight}{.37\textheight}
  \scriptsize
  \centering
  \includegraphics{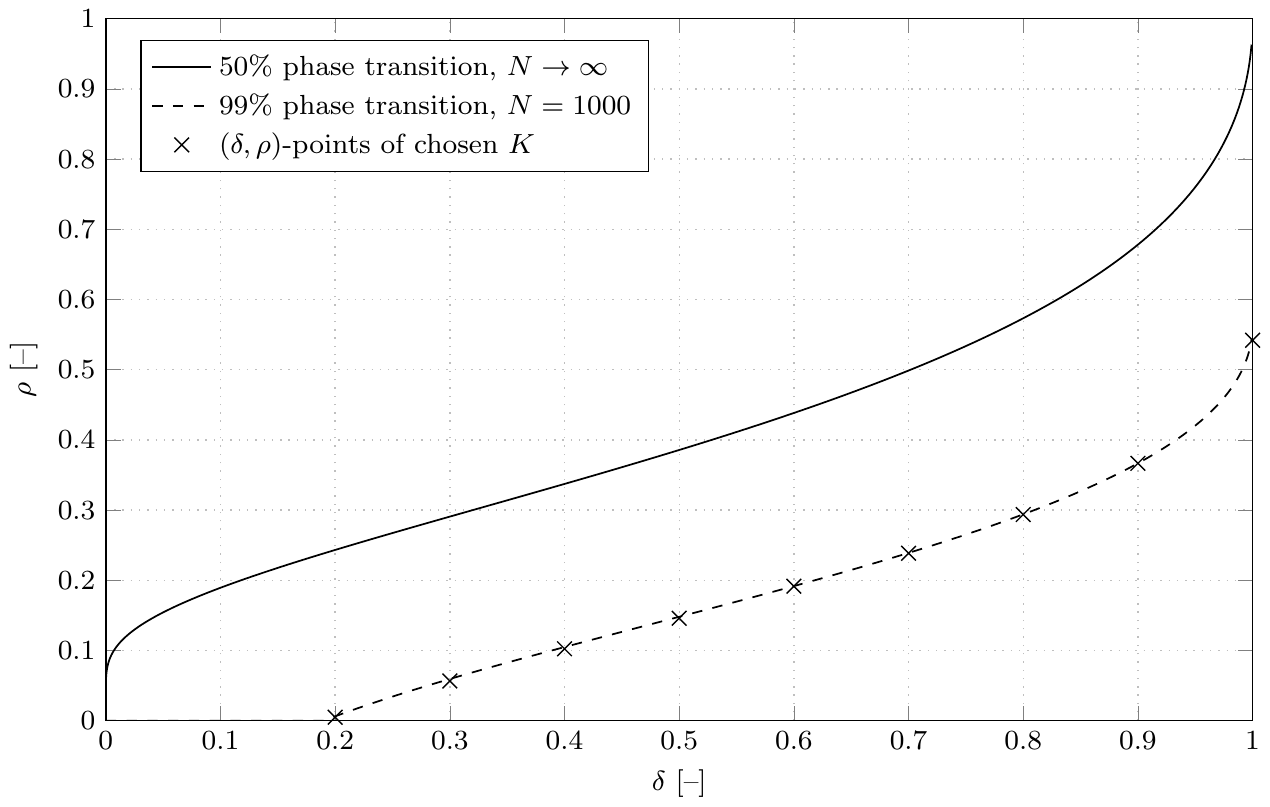}\\
  \caption{The theoretical Donoho-Tanner phase transition for
    $N\rightarrow\infty$, lower bound for $N=1000$, and points
    corresponding to the values of $K$ chosen for test signals.}
  \label{fig:phase-trans-etc}
\end{figure}

Donoho \& Tanner have shown in \cite{Donoho2010} that compressed
sensing problems can be divided into two ``phases'' according to their
probability of correct recovery by the method
(\ref {eq:cvx-sparse-sol}). When evaluating the probability of correct
reconstruction of a sparse vector \cssparsevec over the parameter
space defined by $\delta = \frac{M}{N} \in [0,1]$ and $\rho =
\frac{K}{M} \in [0,1]$, a given problem can be proven to fall into one
of two phases where the probability of correct reconstruction is close
to 1 (feasible) and 0 (infeasible), respectively. These two phases are
divided by a sharp phase transition around the correct reconstruction
probability of $50\%$ as drawn in Fig.~\ref {fig:phase-trans-etc}
(---). The feasible phase lies below the transition and the infeasible
phase lies above. Compressed sensing is utilized most efficiently when
operating close to the phase transition in the feasible phase since
\cssparsevec can be reconstructed with the highest possible number of
non-zero elements $K$, given $N$ and $M$, here. This phase transition
occurs in the case of noiseless measurements, in the limit of
$N\rightarrow\infty$. The theory still holds for finite $N$, but the
phase transition is shifted downwards with respect to $\rho$ in the
$(\delta,\rho)$-parameter space, see Fig.~\ref {fig:phase-trans-etc}
(\mbox{- - -}). It has also been shown that a similar transition
occurs at the same location in the noisy case,
i.e. (\ref {eq:basic-cs-meas})~\cite{Donoho2011}. In the noisy case,
mean squared reconstruction error, $\expectation{\|\cssparsevecest -
\cssparsevec\|_2^2/N}$ relative to the measurement noise variance
$\sigma_\noisesymb^2$ is bounded in the feasible region and unbounded
in the infeasible region.

In all simulations, we apply the proposed approach
to test signals generated
randomly according to the following specifications: size of
\cssparsevec vector $N=1000$; number of compressed measurements
$M\in\{200,300,400,500,600,700,800,900,1000\}$. The non-zero elements
of \cssparsevec are \gls{IID} $\sim\mathcal N(0,1)$; the number of
non-zero elements $K$ is selected for each value of $M$. This is done
by calculating the largest possible $K$ for each $M$ according to the
lower bound on the $99\%$ phase transition for finite $N=1000$ by the
formula given in~\cite[Sec. IV, Theorem 2]{Donoho2010}, drawn in
Fig.~\ref {fig:phase-trans-etc} (\mbox{- - -}). The resulting values are
$K\in\{1,17,41,73,115,167,235,330,542\}$. The corresponding
$(\delta,\rho)$-points are plotted in Fig.~\ref {fig:phase-trans-etc}
($\times$).

The measurement matrix \csmeasmtx has \gls{IID} entries $\sim\mathcal
N(0,\frac{1}{M})$ and we use the dictionary $\csdict = \mathbf I$, so
that $\cssysmtx = \csmeasmtx$. We repeat the experiment $T =
\num{1000}$ times for randomly generated \cssparsevec and \csmeasmtx
in each repetition and average the reconstructed signal \gls{NMSE},
$\mathcal P$, over all solution instances $\cssparsevecest[i],\ i \in
\{1,\ldots,T\}$:
\begin{equation}
  \label{eq:nmse-calc}
  \mathcal P = \frac{1}{T}
  \sum_{i=1}^T
  \frac{\|\cssparsevecest[i]-\cssparsevec[i]\|_2^2}{\|\cssparsevec[i]\|_2^2}.
\end{equation} 
To enable assessment of the quality of the obtained results, we plot
the simulated figures with error bars signifying their $99\%$
confidence intervals computed under the assumption of a Gaussian
distributed mean of the \gls{NMSE}, see e.g.
\cite[Sec. 7.3.1]{Ross2000}. %The error bars are barely visible in the
%figures since they are very tight.
The simulations were conducted for
reconstruction using regular \gls{BPDN} (\ref {eq:cvx-sparse-sol})
vs. our proposed approach (\ref {eq:bpdn-postscale}) (denoted
``BPDN-scale'' in result plots). The numerical
optimization problems were solved using the SPGL1\footnote{SPGL1: A
  solver for large-scale sparse reconstruction
  (\url{http://www.cs.ubc.ca/labs/scl/spgl1}).} software package
\cite{BergFriedlander:2008}.

Regarding the choice of $\epsilon$, for regular \gls{BPDN}
(\ref {eq:cvx-sparse-sol}), we chose $\epsilon$ according to
(\ref {eq:epsilon-choice-becker}), with $\sigma =
\sqrt{\sigma^2_\qnoisesymb}$ from (\ref {eq:q-noise-var}). For our
proposed approach (\ref {eq:bpdn-postscale}), we chose $\epsilon$
according to (\ref {eq:epsilon-choice-becker}), with $\sigma =
\sqrt{\sigma^2_\rnoisesymb}$ from (\ref {eq:r-noise-var}). For both
compared approaches, we consider $\sigma^2_{\bar\csmeassymb}$
known. As demonstrated in Section~\ref {sec:scale-and-reg-values}, $\epsilon$
could be chosen better from empirical observations to provide smaller
error in the reconstruction,
i.e. $\|\cssparsevecest-\cssparsevec\|$. We chose the values
(\ref {eq:epsilon-choice-becker}) as practically useful values for
fairness of the evaluation of our proposed method.

As we have chosen low-rate scalar quantization to demonstrate the
proposed approach to noise correlated with the measurements, we
additionally performed simulations to compare the proposed method to a
state-of-the-art reconstruction algorithm for 1-bit compressed
sensing, \gls{BIHT} \cite{Jacques2013}. This simulation was performed
by evaluating both our proposed method and \gls{BIHT} over the phase
space $\delta,\rho \in [0,1]$ where we discretized the range [0,1] in
steps of $0.01$ for both $\delta$ and $\rho$. In each point
$(\delta,\rho)$ we evaluated $\mathcal P$ according to
(\ref {eq:nmse-calc}) over $T=1000$ repetitions with different
\cssparsevec and \cssysmtx in each instance. For each value $\delta
\in \{0.01,0.02,\ldots,1\}$ we evaluate each of the methods from
$\rho=0.01$ until $\mathcal P > 1$. For \gls{BIHT}, we generated
sparse signals \cssparsevec normalized to $\|\cssparsevec\|_2 = 1$
which is assumed by \gls{BIHT} and other 1-bit compressed sensing
reconstruction algorithms in general. In \gls{BIHT}, estimates
\cssparsevecest are re-normalised after reconstruction which is not
the case in our proposed method.

All scripts required to reproduce the simulation results are openly
accessible\footnote{\url{http://github.com/ThomasA/cs-correlated-noise}}.

\section{Numerical Simulation Results}
\label{sec:numer-sim-results}

In this section we present results of the numerical simulations
conducted according to Section~\ref {sec:simulation-results}.  Firstly, we
evaluate the proposed method under artificial correlated measurement
noise generated according to (\ref {eq:correlation-linear}). Secondly,
we evaluate the method under correlated measurement noise incurred by
scalar quantization of the compressed measurements. These results are
shown in Section~\ref {sec:main-results}. Furthermore, in
Section~\ref {sec:comparison-biht} we present results of simulations comparing
the proposed method to \gls{BIHT}. Finally, in
Section~\ref {sec:scale-and-reg-values} we present results of simulations to
shed light on how the choices of the parameters $\beta$ and $\epsilon$
in (\ref {eq:bpdn-modified-beta}) affect the main results.

\subsection{Main Results}
\label{sec:main-results}

\begin{figure}[!t]
  \setlength{\figurewidth}{.85\columnwidth}
  \setlength{\figureheight}{.2\textheight}
  \scriptsize
  \centering
  \subfloat[Noise of var. equivalent to \SI{1}{\bpsamp}
  quantizer.]{\includegraphics{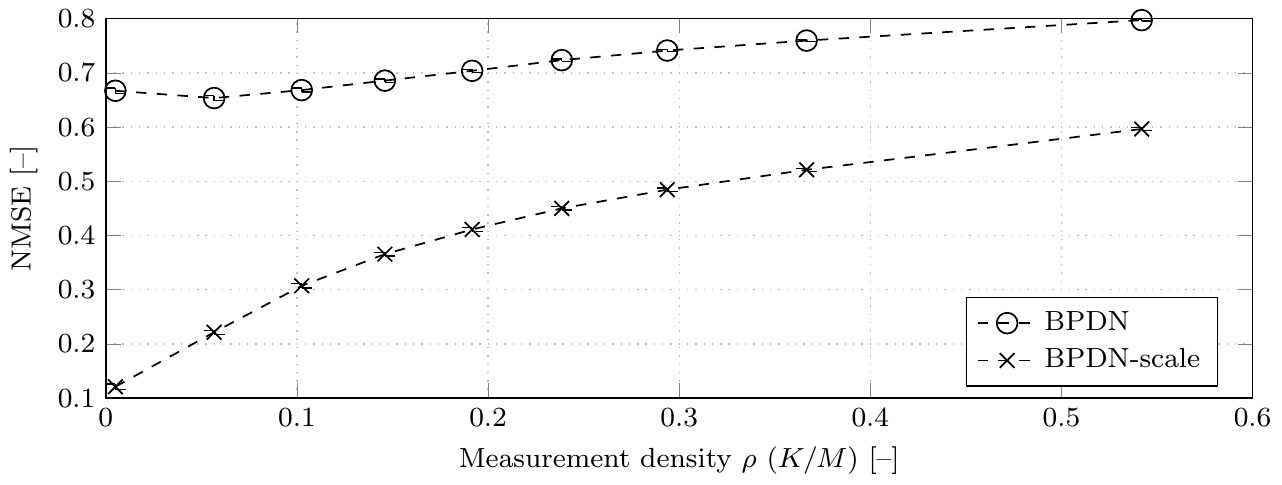}\label{fig:nmse_bpdn_ord_vs_scale_r1}}\\
  \subfloat[Noise of var. equivalent to \SI{3}{\bpsamp}
  quantizer.]{\includegraphics{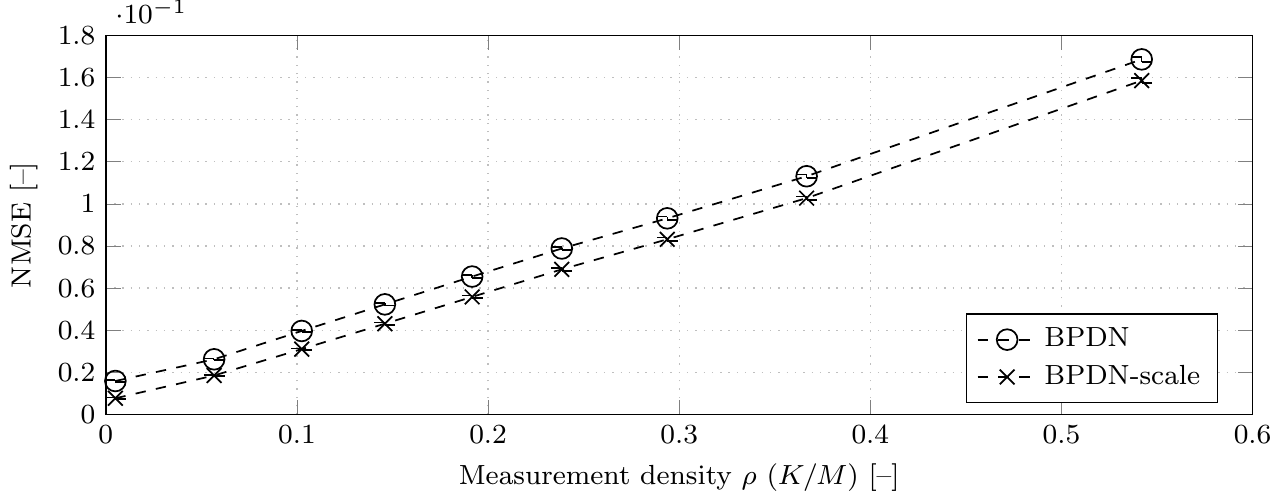}\label{fig:nmse_bpdn_ord_vs_scale_r3}}\\
  \subfloat[Noise of var. equivalent to \SI{5}{\bpsamp}
  quantizer.]{\includegraphics{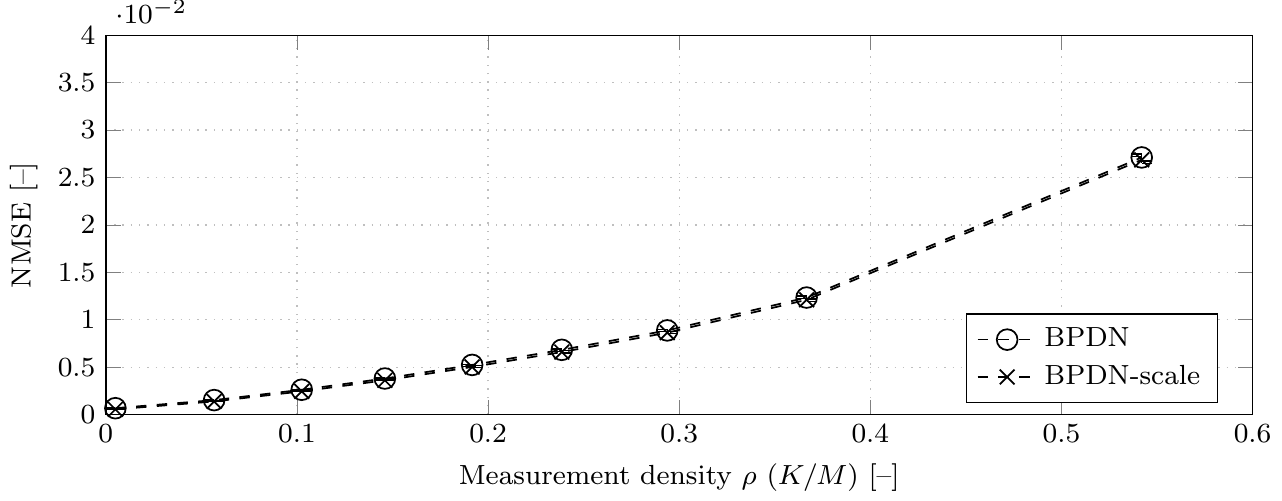}\label{fig:nmse_bpdn_ord_vs_scale_r5}}
  \caption{Simulated NMSE of reconstruction using BPDN vs. relative
    number of measurements for parameters $\alpha$ and
    $\sigma_{\rnoise}^2$ equal to corresponding values for
    Lloyd-Max quantizers.}
  \label{fig:nmse_bpdn_ord_vs_scale}
\end{figure}

In this section, noise variance and correlation parameters are first
set equal to the corresponding parameters estimated for the Lloyd-Max
quantizer used later in this section, for comparability. The parameter
values for $\alpha$ are listed in Table~\ref {tab:alpha-val-bpdn}.
\begin{table}[!t]
  \sisetup{round-mode = places, round-precision = 4}
  \centering
  \caption{Correlation parameter values used in
    Figs.~\ref {fig:nmse_bpdn_ord_vs_scale}--\ref {fig:nmse_bpdn_uni_ord_vs_scale}.}
  \begin{tabular}{rSS}
    Equiv. quantizer resolution & {$\alpha$ (Lloyd-Max)} & {$\alpha$ (uniform)} \\
    \midrule
    \SI{1}{\bpsamp} & 0.636597595 & 0.636597595\\
    \SI{3}{\bpsamp} & 0.965461586 & 0.962583611\\
    \SI{5}{\bpsamp} & 0.997494182 & 0.996506750
  \end{tabular}
  \label{tab:alpha-val-bpdn}
\end{table}
The listed values of $\alpha$ (Lloyd-Max) are used together with
$\sigma_{\rnoise}^2$ calculated from (\ref {eq:r-noise-var}) to generate
correlated measurement noise according to
(\ref {eq:correlation-linear}). In the conducted simulations, \gls{BPDN}
is used to reconstruct \cssparsevecest[2] from the compressed
measurements \csmeas.  We compare the standard (correlation-unaware)
reconstruction, (\ref {eq:cvx-sparse-sol}), of the signal (denoted
``BPDN'' in Fig.~\ref {fig:nmse_bpdn_ord_vs_scale}) to the reconstruction
obtained by our proposed method, (\ref {eq:bpdn-postscale}), of scaling
the reconstructed signal to account for correlation (denoted
``BPDN-scale'' in Fig.~\ref {fig:nmse_bpdn_ord_vs_scale}).  Selected
results for equivalent quantizer resolutions \SI{1}{\bpsamp},
\SI{3}{\bpsamp}, and \SI{5}{\bpsamp} are shown in
Fig.~\ref {fig:nmse_bpdn_ord_vs_scale}.  The proposed method is observed to
improve the reconstruction error $\mathcal P$ by
\SIrange{7.258935064559580}{1.265073250845205}{\deci\bel} (for
increasing $\rho$) at \SI{1}{\bpsamp},
\SIrange{3.078548916636066}{0.259461209546714}{\deci\bel} (for
increasing $\rho$) at \SI{3}{\bpsamp}, and
\SIrange{0.863954374872776}{0.058916440237851}{\deci\bel} (for
increasing $\rho$) at \SI{5}{\bpsamp}.

The experiments for quantized measurements are conducted exactly as
above, with the exception that the measurements \csmeas are quantized
using a \emph{Lloyd-Max
  quantizer}~\cite{Max1960,Lloyd1982}. %or \emph{uniform
% quantizer}~\cite{Bucklew1980} designed for the (Gaussian)
% probability distribution of \csmeas.
The Lloyd-Max quantizer is designed for the Gaussian distribution of
the entries of $\bar\csmeas$ which results from the use of a
measurement matrix containing i.i.d. zero-mean Gaussian entries.
The correlated noise model uses the values of $\alpha$ (Lloyd-Max) for
the selected quantizer resolutions listed
in~Table~\ref {tab:alpha-val-bpdn}.

\begin{figure}[!t]
  \setlength{\figurewidth}{.85\columnwidth}
  \setlength{\figureheight}{.2\textheight}
  \scriptsize
  \centering
  \subfloat[\SI{1}{\bpsamp}
  quantizer.]{\includegraphics{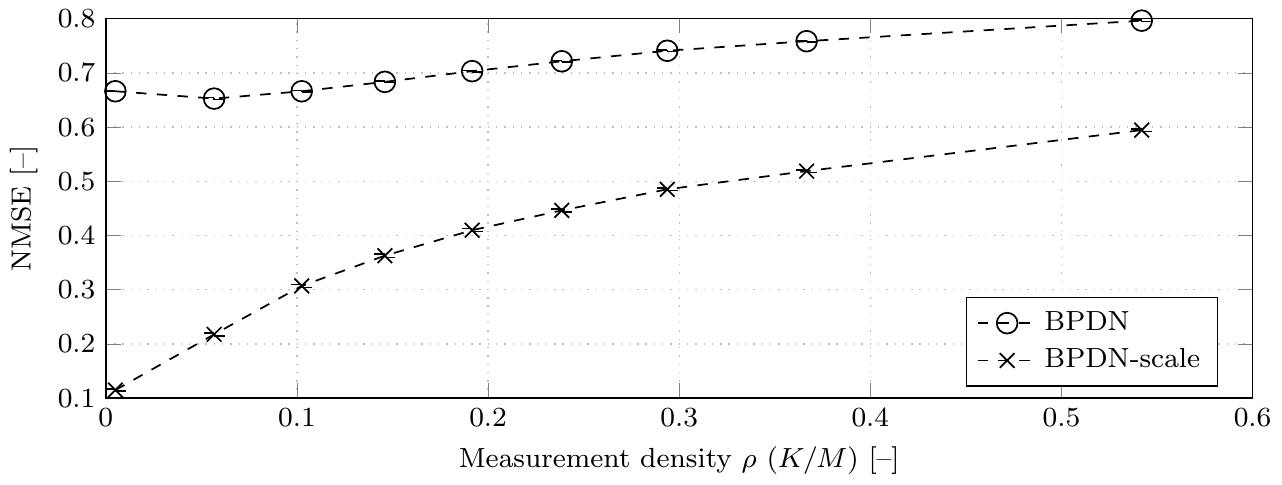}\label{fig:nmse_bpdn_lm_ord_vs_scale_r1}}\\
  \subfloat[\SI{3}{\bpsamp}
  quantizer.]{\includegraphics{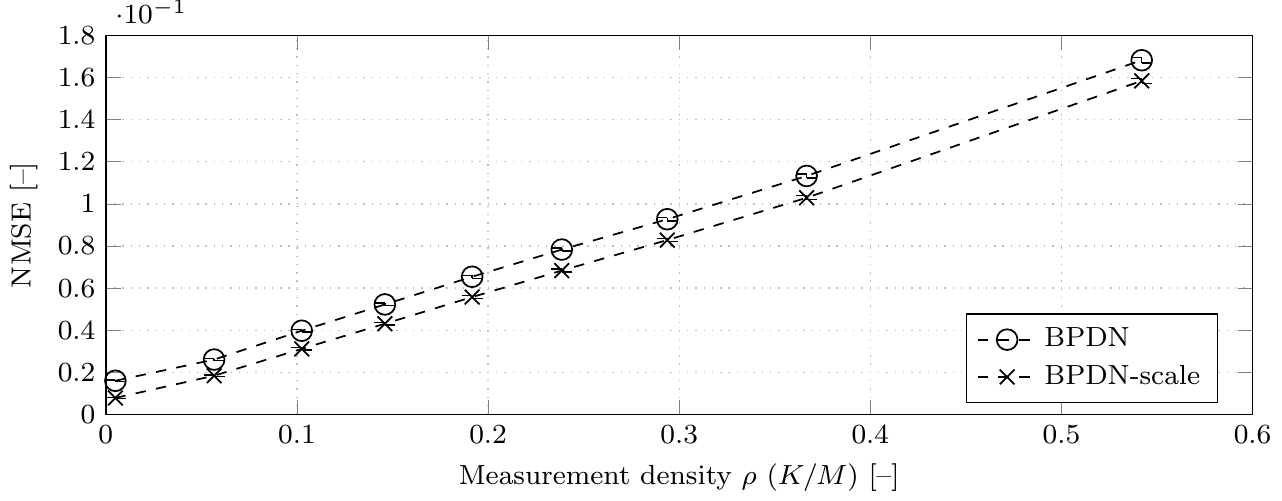}\label{fig:nmse_bpdn_lm_ord_vs_scale_r3}}\\
  \subfloat[\SI{5}{\bpsamp}
  quantizer.]{\includegraphics{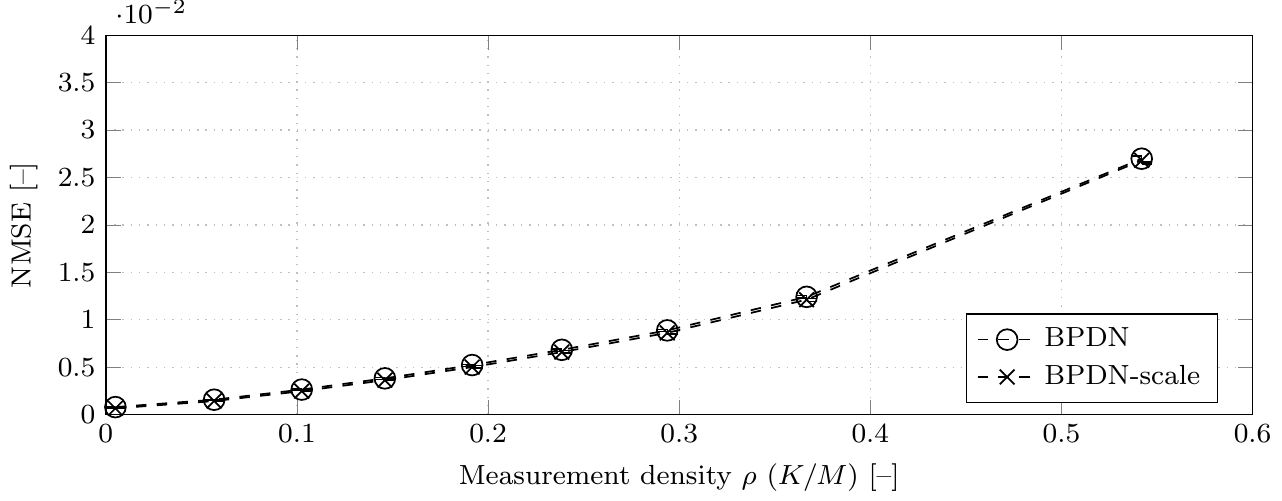}\label{fig:nmse_bpdn_lm_ord_vs_scale_r5}}
  \caption{Simulated NMSE of reconstruction using BPDN vs. relative
    number of measurements for parameters $\alpha$ and
    $\sigma_{\rnoise}^2$ equal to corresponding values for
    Lloyd-Max quantizers.}
  \label{fig:nmse_bpdn_lm_ord_vs_scale}
\end{figure}

Selected results for quantizer resolutions \SI{1}{\bpsamp},
\SI{3}{\bpsamp}, and \SI{5}{\bpsamp} with Lloyd-Max quantization are
shown in Fig.~\ref {fig:nmse_bpdn_lm_ord_vs_scale}. It can be observed that
the reconstruction error figures $\mathcal P$ agree well with those
simulated with artificially generated correlated noise in
Fig.~\ref {fig:nmse_bpdn_ord_vs_scale}. The observed improvements by the
proposed method are almost identical to those observed for artificial
noise: \SIrange{7.630424592316444}{1.269499068230857}{\deci\bel} at
\SI{1}{\bpsamp},
\SIrange{3.069402388553215}{0.260195090772186}{\deci\bel} at
\SI{3}{\bpsamp}, and
\SIrange{0.795311069099512}{0.027794290505028}{\deci\bel} at
\SI{5}{\bpsamp}.

To evaluate our proposed approach for a more practical quantization
scheme than the non-uniform Lloyd-Max quantizer, we additionally
simulated results where the measurements \csmeas are quantized using a
uniform quantizer with mid-point quantization points, optimized for
\gls{MMSE} of the quantized measurements.
The uniform quantizer is designed for the Gaussian distribution of
the entries of $\bar\csmeas$.
This serves to evaluate how
well the proposed approach performs for a more practical quantizer
type that does not theoretically obey the quantization noise model
(\ref {eq:gain-plus-additive-q}) due to the fact that its reconstruction
points are not the centroids of the input signal's \gls{PDF} in the
quatizer's input regions. The correlated noise model uses the values
of $\alpha$ (uniform) from Table~\ref {tab:alpha-val-bpdn}.

\begin{figure}[!t]
  \setlength{\figurewidth}{.85\columnwidth}
  \setlength{\figureheight}{.2\textheight}
  \scriptsize
  \centering
  \subfloat[\SI{1}{\bpsamp}
  quantizer.]{\includegraphics{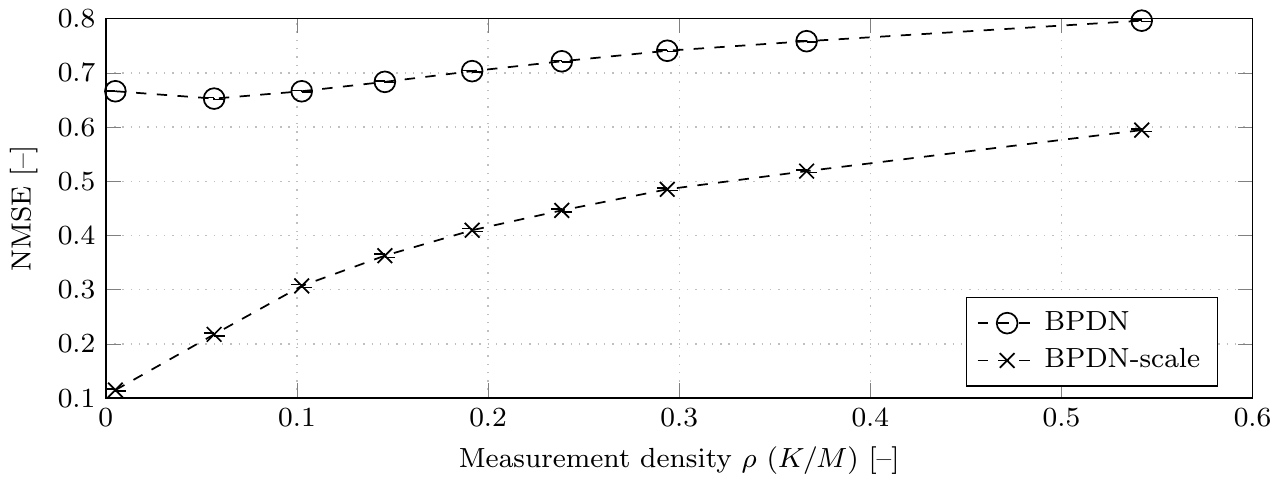}\label{fig:nmse_bpdn_uni_ord_vs_scale_r1}}\\
  \subfloat[\SI{3}{\bpsamp}
  quantizer.]{\includegraphics{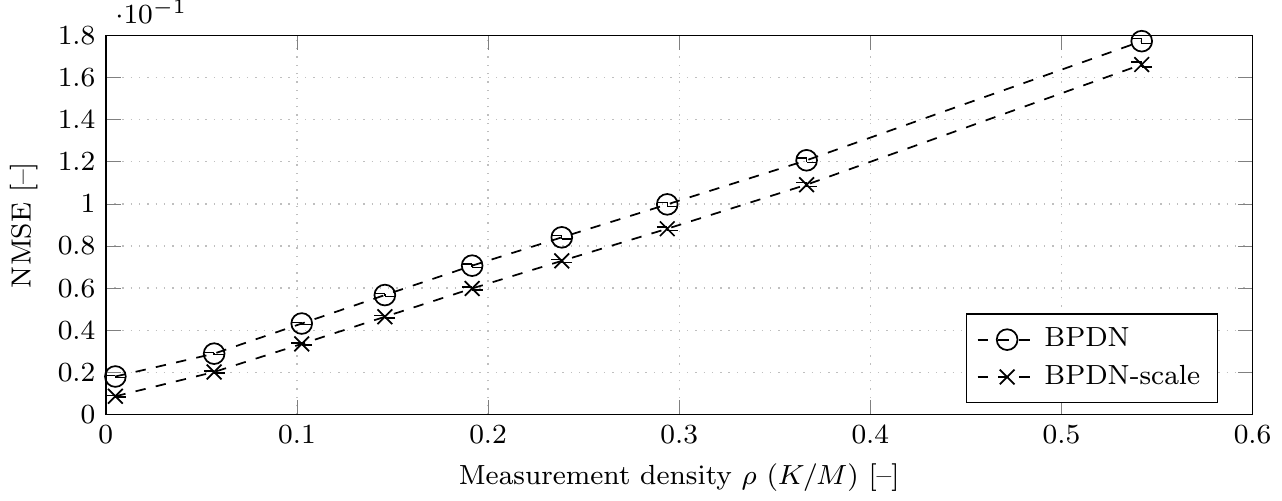}\label{fig:nmse_bpdn_uni_ord_vs_scale_r3}}\\
  \subfloat[\SI{5}{\bpsamp}
  quantizer.]{\includegraphics{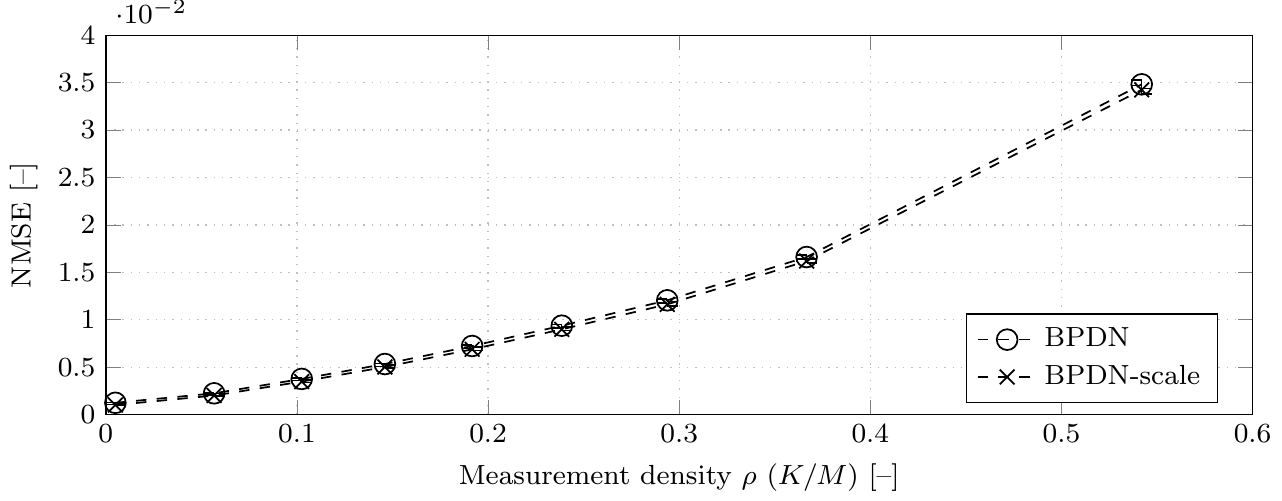}\label{fig:nmse_bpdn_uni_ord_vs_scale_r5}}
  \caption{Simulated NMSE of reconstruction using BPDN vs. relative
    number of measurements for parameters $\alpha$ and
    $\sigma_{\rnoise}^2$ equal to corresponding values for uniform
    quantizers.}
  \label{fig:nmse_bpdn_uni_ord_vs_scale}
\end{figure}

Selected results for quantizer resolutions \SI{1}{\bpsamp},
\SI{3}{\bpsamp}, and \SI{5}{\bpsamp} with uniform quantization are
shown in Fig.~\ref {fig:nmse_bpdn_uni_ord_vs_scale}. The observed
improvements by the proposed method are close to those observed for
artificial noise:
\SIrange{7.630376055876695}{1.269499006131298}{\deci\bel} at
\SI{1}{\bpsamp},
\SIrange{3.171612435612246}{0.280326866601740}{\deci\bel} at
\SI{3}{\bpsamp}, and
\SIrange{0.892202647533395}{0.072544852438341}{\deci\bel} at
\SI{5}{\bpsamp}.
The results in
Fig.~\ref {fig:nmse_bpdn_lm_ord_vs_scale_r1} and~\ref {fig:nmse_bpdn_uni_ord_vs_scale_r1}
are identical due to the fact that the 2-level Lloyd-Max quantizer is
a uniform 2-level quantizer optimized for \gls{MMSE} of the quantized
values. It can also be observed that the uniform quantizer for
\SIlist{3;5}{\bpsamp} results in slightly larger reconstruction error
while the improvement by our proposed method is preserved.

\subsection{Comparison to \acrfull{BIHT}}
\label{sec:comparison-biht}

\begin{figure}[!t]
  \setlength{\figurewidth}{.85\columnwidth}
  \setlength{\figureheight}{.37\textheight}
  \scriptsize
  \centering
  \subfloat[Proposed method
  (BPDN-scale).]{\includegraphics{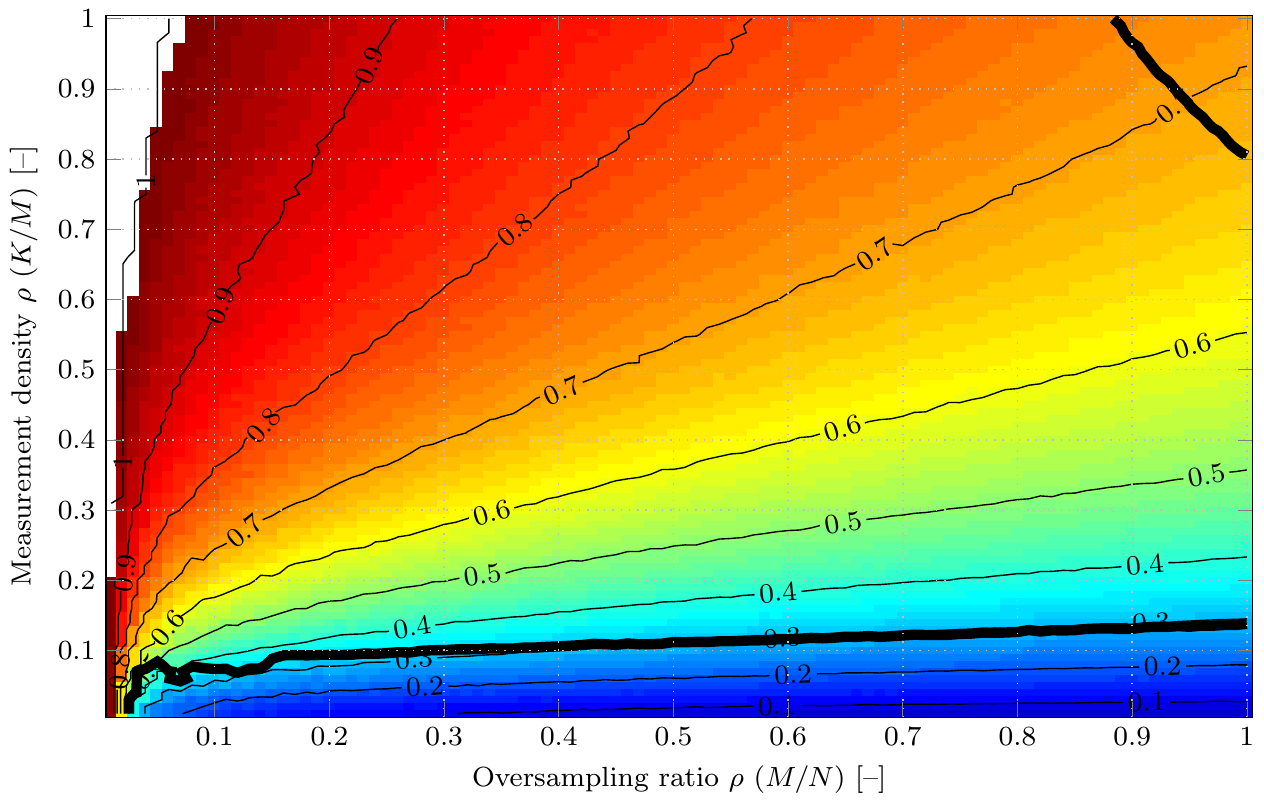}\label{fig:nmse_bpdn-scale}}\\
  \subfloat[BIHT.]{\includegraphics{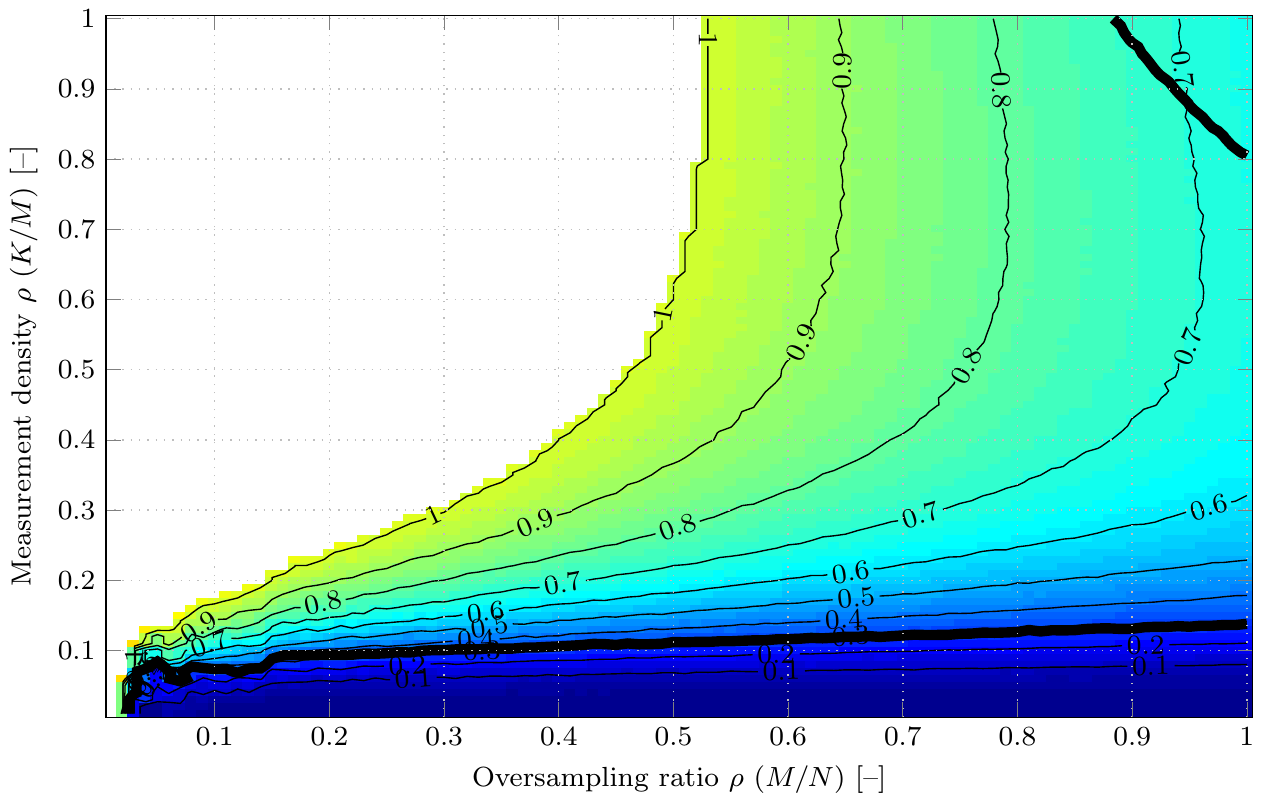}\label{fig:nmse_biht}}
  \caption{Simulated NMSE of reconstruction from 1-bit quantized
    measurements. The numbered (---$0.1$--- etc.) contour lines trace
    equal NMSE levels. The bold contour lines (\textbf{---}) mark the
    boundary where the NMSE levels of the proposed method and BIHT,
    respectively, are equal. (``Wiggly'' contour lines are caused by
    interpolation in Matlab).}
  \label{fig:nmse_bpdn-scale_vs_biht}
\end{figure}

In this section, we provide results comparing our proposed method to
\gls{BIHT}. Results for our proposed method were computed in the same
manner as for the results regarding 1-bit quantization in
Section~\ref {sec:main-results}. The simulated \gls{NMSE} of our proposed
method and \gls{BIHT} are shown in
Fig.~\ref {fig:nmse_bpdn-scale_vs_biht}. The white regions of the phase
space are un-tested as they lie beyond $\mathcal P > 1$; a threshold
we selected to define the region we wished to investigate. The bold
contour lines mark the boundary where the \glspl{NMSE} of our
proposed method and \gls{BIHT} are equal. As the numbered contour
lines show, ``BPDN-scale'' exhibits lower \gls{NMSE} than \gls{BIHT}
in the majority (upper left region) of the phase space, whereas the
\gls{NMSE} of \gls{BIHT} is lower along the bottom of the phase
space -- up to around $\rho=0.1$ -- and in the upper right-hand
corner -- towards $(\delta,\rho)=(1,1)$.

\subsection{Empirical Investigation of Scaling Factors and
  Regularization Parameters}
\label{sec:scale-and-reg-values}

In order to assess the optimality of the proposed approach as
described in Section~\ref {sec:optim-prop-appr}, we conducted simulations for
values of $\beta$ in (\ref {eq:bpdn-modified-beta}) using artificial
pseudo-random noise generated according to the model
(\ref {eq:gain-plus-additive-q}).  Since the reconstruction error
performance is also affected by the choice of $\epsilon$ in
(\ref {eq:bpdn-modified-beta}), we similarly performed the simulations
over different values $\epsilon$. Preliminary simulations indicated
that $\mathcal P$ (see (\ref {eq:nmse-calc})) evolves in a quasi-convex
manner over $\beta$ and $\epsilon$. Based on this observation, we have
used the Nelder-Mead simplex algorithm \cite{Nelder1965} to find the
$(\beta, \epsilon)$-optimal error figures $\mathcal P$ for each of the
points $(M,K)$ listed in Section~\ref {sec:simulation-results}. The results
for all $(M,K)$ with correlated noise generated according to each of
the values $\alpha$ (Lloyd-Max) in Table~\ref {tab:alpha-val-bpdn} are shown
in tables~ \ref {tab:opt-nmse-1} to~ \ref {tab:opt-nmse-5} . The optimal regularization parameter values
for ordinary \gls{BPDN} are denoted $\epsilon_1$ --  with resulting
error figure $\mathcal P_1$, while the optimal
scaling and regularization parameter values for the proposed method
are denoted $\beta_2$ and $\epsilon_2$ -- with resulting
error figure $\mathcal P_2$. The error figures from our proposed
method as reported in Fig.~\ref {fig:nmse_bpdn_ord_vs_scale} are included
in tables~ \ref {tab:opt-nmse-1} to~ \ref {tab:opt-nmse-5}  as $\mathcal
P_\alpha$ to facilitate comparison.
\begin{table}[!t]
  \sisetup{round-mode = figures,
    round-precision = 2,
    %output-exponent-marker = \text{e},
    exponent-product = \cdot,
    tight-spacing = true
  }
  \centering
  \caption{Simulated NMSE at empirically optimal parameter values
    $\beta$ and $\epsilon$. Noise equivalent to \SI{1}{\bpsamp}
    quantizer.}
  \begin{tabular}{rSS[table-number-alignment=left,
      scientific-notation=true,
      table-figures-exponent=1]SSS[table-number-alignment=left,
      scientific-notation=true, table-figures-exponent=1]S[table-number-alignment
      = left,scientific-notation=true,table-figures-exponent=1]}
    $(M,K)$ & {$\epsilon_1/\epsilon$} & {$\mathcal{P}_1$} &
    {$\beta_2/\alpha$} & {$\epsilon_2/\epsilon$} &
    {$\mathcal{P}_2$} & {$\mathcal{P}_\alpha$}\\
    \midrule
    $(200, 1)$ & 0.600000094 & 1.962764128e-02 & 0.836230469 & 0.921484375 & 1.019192831e-03    & 1.211786045e-01\\
    $(300, 17)$ & 0.499999905 & 3.309425673e-02 & 0.738657352 & 0.925879428 & 1.366921566e-02   & 2.215994094e-01\\
    $(400, 41)$ & 0.512988281 & 3.888425539e-02 & 0.829915241 & 0.865297454 & 2.248581511e-02   & 3.073345237e-01\\
    $(500, 73)$ & 0.4625 & 4.271101597e-02 & 0.772141054 & 0.931689703 & 2.889661903e-02        & 3.657011687e-01\\
    $(600, 115)$ & 0.512597632 & 4.472722152e-02 & 0.792977346 & 0.90156264 & 3.445824343e-02   & 4.112594208e-01\\
    $(700, 167)$ & 0.474993896 & 4.605901923e-02 & 0.882593934 & 0.843647052 & 3.801668597e-02  & 4.501421697e-01\\
    $(800, 235)$ & 0.499951172 & 4.783124560e-02 & 0.997998047 & 0.739648437 & 4.172978167e-02  & 4.846145118e-01\\
    $(900, 330)$ & 0.456152368 & 4.895679800e-02 & 1.03492432 & 0.730126953 & 4.527469529e-02   & 5.212606004e-01\\
    $(1000, 542)$ & 0.463378906 & 5.265562631e-02 & 1.15138943 & 0.654899013 & 5.122387807e-02  & 5.965638647e-01\\
  \end{tabular}
  \label{tab:opt-nmse-1}
\end{table}
\begin{table}[!t]
  \sisetup{round-mode = figures,
    round-precision = 2,
    %output-exponent-marker = \text{e},
    exponent-product = \cdot,
    tight-spacing = true
  }
  \centering
  \caption{Simulated NMSE at empirically optimal parameter values
    $\beta$ and $\epsilon$. Noise equivalent to \SI{3}{\bpsamp}
    quantizer.}
  \begin{tabular}{rSS[table-number-alignment =
      left,scientific-notation=true,table-figures-exponent=1]SSS[table-number-alignment
      = left,scientific-notation=true,table-figures-exponent=1]S[table-number-alignment
      = left,scientific-notation=true,table-figures-exponent=1]}
    $(M,K)$ & {$\epsilon_1/\epsilon$} & {$\mathcal{P}_1$} &
    {$\beta_2/\alpha$} & {$\epsilon_2/\epsilon$} &
    {$\mathcal{P}_2$} & {$\mathcal{P}_\alpha$}\\
    \midrule
    $(200, 1)$ & 0.8359375 & 5.003686966e-04 & 0.895251465 & 1.05541382 & 3.845094096e-05      & 7.777292419e-03\\ 
    $(300, 17)$ & 0.7515625 & 1.956225894e-03 & 0.914624023 & 0.929260254 & 8.878818358e-04    & 1.858795845e-02\\ 
    $(400, 41)$ & 0.639453125 & 3.003703742e-03 & 0.883948517 & 0.962291336 & 1.799853026e-03  & 3.110813475e-02\\ 
    $(500, 73)$ & 0.646875 & 3.925235888e-03 & 0.870380402 & 0.92620182 & 2.706928466e-03      & 4.315267758e-02\\ 
    $(600, 115)$ & 0.65 & 4.805505342e-03 & 0.921708177 & 0.737433008 & 3.578545485e-03        & 5.587304139e-02\\ 
    $(700, 167)$ & 0.6109375 & 5.639735722e-03 & 0.904189301 & 0.730999565 & 4.454644609e-03   & 6.893618006e-02\\ 
    $(800, 235)$ & 0.549609375 & 6.412675786e-03 & 0.915832612 & 0.674575924 & 5.402928323e-03 & 8.318545949e-02\\ 
    $(900, 330)$ & 0.500195312 & 7.433060562e-03 & 0.92930154 & 0.627553277 & 6.762944552e-03  & 1.027183318e-01\\ 
    $(1000, 542)$ & 0.4875 & 1.044461637e-02 & 0.981768556 & 0.505537388 & 1.023121716e-02     & 1.584817387e-01\\ 
  \end{tabular}
  \label{tab:opt-nmse-3}
\end{table}
\begin{table}[!t]
  \sisetup{round-mode = figures,
    round-precision = 2,
    %output-exponent-marker = \text{e},
    exponent-product = \cdot,
    tight-spacing = true
  }
  \centering
  \caption{Simulated NMSE at empirically optimal parameter values
    $\beta$ and $\epsilon$. Noise equivalent to \SI{5}{\bpsamp}
    quantizer.}
  \begin{tabular}{rSS[table-number-alignment =
      left,scientific-notation=true,table-figures-exponent=1]SSS[table-number-alignment
      = left,scientific-notation=true,table-figures-exponent=1]S[table-number-alignment
      = left,scientific-notation=true,table-figures-exponent=1]}
    $(M,K)$ & {$\epsilon_1/\epsilon$} & {$\mathcal{P}_1$} &
    {$\beta_2/\alpha$} & {$\epsilon_2/\epsilon$} &
    {$\mathcal{P}_2$} & {$\mathcal{P}_\alpha$}\\
    \midrule
    $(200, 1)$ & 0.831152344 & 1.799160130e-05 & 0.972290802 & 1.03970718 & 2.574893847e-06    & 5.589140078e-04\\
    $(300, 17)$ & 0.69921875 & 1.181667478e-04 & 0.973400879 & 0.935858154 & 6.284629468e-05   & 1.395401644e-03\\
    $(400, 41)$ & 0.6546875 & 2.009662307e-04 & 0.967785645 & 0.961956787 & 1.295546072e-04    & 2.457189756e-03\\
    $(500, 73)$ & 0.68125 & 2.827824401e-04 & 0.970315552 & 0.846421814 & 1.904358375e-04      & 3.617261602e-03\\
    $(600, 115)$ & 0.631445312 & 3.756479372e-04 & 0.968706578 & 0.723657039 & 2.740219755e-04 & 5.028247464e-03\\
    $(700, 167)$ & 0.5 & 4.768401725e-04 & 0.964893913 & 0.721007252 & 3.628962904e-04         & 6.600790772e-03\\
    $(800, 235)$ & 0.56328125 & 5.768559384e-04 & 0.969677734 & 0.659985352 & 4.776633846e-04  & 8.639184807e-03\\
    $(900, 330)$ & 0.4484375 & 7.326732124e-04 & 0.964321676 & 0.575938128 & 6.383322100e-04   & 1.210708841e-02\\
    $(1000, 542)$ & 0.4140625 & 1.315488557e-03 & 0.980423737 & 0.373928452 & 1.255090752e-03  & 2.686228056e-02\\
  \end{tabular}
  \label{tab:opt-nmse-5}
\end{table}

It was expected that $\alpha$ would be the optimal choice of $\beta$,
i.e. $\beta = \alpha$.  However, it turns out that the (empirically
observed) optimal value of $\beta_2$ is typically slightly smaller
than $\alpha$ with observed values $\beta_2 \in
[\num{0.738657352}\alpha, \num{0.981768556}\alpha]$,
depending on $(M,K)$. An exception is seen in Table~\ref {tab:opt-nmse-1},
where $\beta_2 \in [\num{1.03492432}\alpha,
\num{1.15138943}\alpha]$.

The optimal values of the regularization parameter $\epsilon$ are
similarly found to be lower than the values given by
(\ref {eq:epsilon-choice-becker}). For the baseline method
(\ref {eq:cvx-sparse-sol}), the (empirically observed) optimal values
are observed as $\epsilon_1 \in [\num{0.4140625}\epsilon,
\num{0.8359375}\epsilon]$, depending on $(M,K)$, where
$\epsilon$ denotes the values given by (\ref {eq:epsilon-choice-becker})
as described in Section~\ref {sec:simulation-results}. For our proposed method
(\ref {eq:bpdn-postscale}), the optimal values are generally closer to
the values given by (\ref {eq:epsilon-choice-becker}) with observed
values $\epsilon_2 \in [\num{0.373928452}\epsilon,
\num{1.05541382}\epsilon]$, depending on $(M,K)$.

It is important to note that the demonstrated advantage of our
proposed approach in Section~\ref {sec:main-results} is not merely a result of
a particularly lucky choice of $\epsilon$, as these experiments
testify. The observed \gls{NMSE} of our proposed method,
$\mathcal{P}_2$, consistently outperforms the baseline approach,
$\mathcal{P}_1$. The improvement is consistent across different
correlation parameters $\alpha$ as seen in tables~ \ref {tab:opt-nmse-1} to~ \ref {tab:opt-nmse-5}  where
$\mathcal{P}_2$ is smaller than $\mathcal{P}_1$ by
\SIlist{12.8461;11.1438;8.4431}{\deci\bel}
in tables~ \ref {tab:opt-nmse-1} to~ \ref {tab:opt-nmse-5} , respectively,
for $(M,K) = (200,1)$. At the other extreme of $(M,K) = (1000,542)$,
$\mathcal{P}_2$ is smaller than $\mathcal{P}_1$ by
\SIlist{0.1197;0.0897;0.2041}{\deci\bel},
respectively. Additionally, the observed \glspl{NMSE} $\mathcal P_2$
are generally around an order of magnitude lower than $\mathcal
P_\alpha$ arising from our proposed choices of $\beta=\alpha$ and
$\epsilon$ according to (\ref {eq:epsilon-choice-becker}). However, note
that $\beta_2$ and $\epsilon_2$ optimized through simulations are not
practically useful.

\section{Discussion}
\label{sec:discussion}

As seen from the experimental results in
Section~\ref {sec:scale-and-reg-values}, the correlation parameter $\alpha$
from (\ref {eq:correlation-linear}) may in fact not be the optimal
choice of scaling parameter, as expressed by $\beta$ in
(\ref {eq:bpdn-modified-beta}). The generally smaller values found in
Section~\ref {sec:scale-and-reg-values} to be optimal for \gls{BPDN}
reconstruction according to (\ref {eq:bpdn-modified-beta}) can be
explained by the fact that they scale the estimate
\cssparsevecest[\beta] larger. It is well-known in the literature that
the $\ell_1$-norm minimization approach represented by, e.g.,
(\ref {eq:cvx-sparse-sol}) tends to penalize larger coefficients of
\cssparsevec more than smaller coefficients \cite{Candes2008a}, thus
estimating the former relatively too small. Therefore, it is possible
to choose a scaling parameter $\beta < \alpha$ in
(\ref {eq:bpdn-modified-beta}) that improves the estimate
\cssparsevecest[\beta], i.e. yields smaller $\|\cssparsevecest[\beta]
- \cssparsevec\|$ compared to $\|\cssparsevecest[\alpha] -
\cssparsevec\|$. At this time, we cannot quantify the optimal $\beta$
analytically and it depends on the indeterminacy and/or measurement
density of the performed compressed sensing.

Regarding the comparison of the proposed method to BIHT, the two
methods require two different kinds of prior information. BIHT
requires knowing that the sparse vector \cssparsevec is unit-norm:
$\|\cssparsevec\|_2 = 1$. Our proposed method requires knowing the
variance of the unquantized measurements $\bar\csmeassymb$ -- the
elements of $\bar\csmeas$. It may depend on the specific application
which quantity is more realistic to know about the signal. At least,
the variance assumed known in our proposed method does not require any
knowledge (such as norm) of the sparse representation \cssparsevec of
the observed signal.

\section{Conclusion}
\label{sec:conclusion}

We proposed a simple technique to model correlation between
measurements and an additive noise in compressed sensing signal
reconstruction. The technique is based on a linear model of the
correlation between the measurements and noise. It consists of scaling
signals reconstructed by a well-known $\ell_1$-norm convex
optimization method according to the model and comes at negligible
computational cost. We provided practical expressions for computing
the scaling parameter and the reconstruction regularization parameter.

We performed numerical simulations to demonstrate the obtainable
reconstruction error improvement by the proposed method compared to
ordinary $\ell_1$-norm convex optimization reconstruction for noise
generated according to the model. We further demonstrated as an
example that the model applies well to low-rate scalar quantization of
the measurements; both Lloyd-Max quantization that complies accurately
with the correlation model, as well as the more practical uniform
quantization. For example, simulations indicated that the proposed
method offers improvements on the order of \SIrange{1}{7}{\deci\bel}
for \SI{1}{\bpsamp} quantization, depending on the indeterminacy of
the performed compressed sensing.

We compared the proposed approach to a state-of-the-art reconstruction
method, \glsfirst{BIHT}, for the special case of \SI{1}{\bpsamp}
quantization. This comparison showed that the proposed
approach reconstructs signals with smaller error than \gls{BIHT} when
the signals contain more non-zero elements than an approximate
fraction of $0.1$ of the number of measurements. This indicated that the
proposed method is able to reconstruct less sparse signals from 1-bit
quantized measurements than \gls{BIHT} is capable of.

We conducted numerical simulations to evaluate the validity of our
results which confirmed that the improvements offered by the proposed
method are not merely a coincidental result of the suggested practical
choices of scaling and optimization regularization parameters. These
results further indicated that the proposed method is robust to the
choice of scaling and optimization regularization parameter in the
sense that a suboptimal choice still leads to considerable
improvements over the ordinary convex optimization reconstruction
method.

\section*{Acknowledgements}

This work was partially financed by The Danish Council for Strategic
Research under grant number 09-067056 and by the Danish Center for
Scientific Computing.

\section*{References}

%\printbibliography
\bibliographystyle{elsarticle-num}
\bibliography{bibliography} % Use "bibtex8 -H", not "bibtex"

\end{document}

%% file: acronyms.tex
\newacronym{AD}{A/D}{analog-to-digital}
\newacronym{AIC}{AIC}{analog-to-information converter}
\newacronym{AIHT}{AIHT}{accelerated iterative hard thresholding}
\newacronym{AR}{AR}{auto-regressive}
\newacronym{ARMA}{ARMA}{auto-regressive moving average}
\newacronym{AWN}{AWN}{additive white noise}
\newacronym{BIHT}{BIHT}{Binary Iterative Hard Thresholding}
\newacronym{BPDN}{BPDN}{Basis Pursuit De-Noising}
\newacronym{CC}{CC}{cross-correlated}
\newacronym{CELP}{CELP}{code-excited linear prediction}
\newacronym{COSAMP}{CoSaMP}{Compressive Sampling Matching Pursuit}
\newacronym{CS}{CS}{compressed sensing}
\newacronym{DA}{D/A}{digital-to-analog}
\newacronym{DFT}{DFT}{discrete Fourier transform}
\newacronym{DHT}{DHT}{discrete Hartley transform}
\newacronym{DPCM}{DPCM}{differential pulse code modulation}
\newacronym{DVBT}{\mbox{DVB-T}}{Digital Video Broadcasting - Terrestrial}
\newacronym{DVBH}{\mbox{DVB-H}}{Digital Video Broadcasting - Handheld}
\newacronym{ECC}{ECC}{error eorrecting code}
\newacronym{EKF}{EKF}{extended Kalman filter}
\newacronym{FEC}{FEC}{forward error correction}
\newacronym{FFT}{FFT}{fast Fourier transform}
\newacronym{FIR}{FIR}{finite impulse response}
\newacronym{GMM}{GMM}{Gaussian mixture model}
\newacronym{HMM}{HMM}{hidden Markov model}
\newacronym{ICASSP}{ICASSP}{IEEE International Conference on Acoustics,
  Speech, and Signal Processing}
\newacronym{IID}{i.i.d.}{independent identically distributed}
\newacronym{IHT}{IHT}{Iterative Hard Thresholding}
\newacronym{JLS}{JLS}{jump linear system}
\newacronym{KF}{KF}{Kalman filter}
\newacronym{LASSO}{LASSO}{Least Absolute Shrinkage and Selection Operator}
\newacronym[firstplural=LMIs (linear matrix inequalities)]{LMI}{LMI}{linear matrix inequality}
\newacronym{LMMSE}{LMMSE}{linear minimum mean-squared error}
\newacronym{LPC}{LPC}{linear predictive coding}
\newacronym[firstplural=LSFs (line spectral frequencies)]{LSF}{LSF}{line spectral frequency}
\newacronym{LTI}{LTI}{linear time-invariant}
\newacronym{MA}{MA}{moving average}
\newacronym{MAC}{MAC}{media access control}
\newacronym{MDC}{MDC}{multiple description coding}
\newacronym[firstplural=MDPs (Markov decision processes)]{MDP}{MDP}{Markov decision process}
\newacronym{MMSE}{MMSE}{minimum mean squared error}
\newacronym{MSE}{MSE}{mean squared error}
\newacronym{MWC}{MWC}{modulated wideband converter}
\newacronym{NMSE}{NMSE}{Normalized Mean Squared Error}
\newacronym{OSI}{OSI}{Open Systems Interconnection}
\newacronym{PDF}{p.d.f.}{probability density function}
\newacronym{QoS}{QoS}{quality of service}
\newacronym{RD}{RD}{random demodulator}
\newacronym{RF}{RF}{radio frequency}
\newacronym{RIP}{RIP}{Restricted Isometry Property}
\newacronym{RTP}{RTP}{Real-time Transport Protocol}
\newacronym{RTCP}{RTCP}{RTP Control Protocol}
\newacronym{SNR}{SNR}{signal-to-noise ratio}
\newacronym{SP}{SP}{Subspace Pursuit}
\newacronym{TCP}{TCP}{Transmission Control Protocol}
\newacronym{TST}{TST}{Two-Stage Thresholding}
\newacronym{UC}{UC}{uncorrelated}
\newacronym{UDP}{UDP}{User Datagram Protocol}
\newacronym{UHF}{UHF}{ultra high frequency}
\newacronym{UKF}{UKF}{unscented Kalman filter}
\newacronym{UUP}{UUP}{uniform uncertainty principle}
\newacronym{UWB}{UWB}{ultra-wideband}
\newacronym{VOIP}{VoIP}{voice over IP}
\newacronym{GE}{GE}{Gilbert-Elliot}
% Acronyms for methods in my paper
\newacronym{IMVI}{IMVI}{Iterative Measurement Vector Improvement}
\newacronym{GVS}{GVS}{Gain Vector Search}

%% file: cs_math.tex
\newcommand{\csmeassymb}{y}
\newcommand{\csmeas}[1][]{\ensuremath{\mathbf\csmeassymb_{#1}}\xspace}
\newcommand{\cssparsevecsymb}{x}
\newcommand{\cssparsevec}[1][]{\ensuremath{\mathbf \cssparsevecsymb_{#1}}\xspace}
\newcommand{\cssparsevecvarsymb}{u}
\newcommand{\cssparsevecvar}[1][]{\ensuremath{\mathbf \cssparsevecvarsymb_{#1}}\xspace}
\newcommand{\cssparsevecvaraltsymb}{v}
\newcommand{\cssparsevecvaralt}[1][]{\ensuremath{\mathbf \cssparsevecvaraltsymb_{#1}}\xspace}
\newcommand{\cssparsevecest}[1][]{\ensuremath{\hat{\mathbf \cssparsevecsymb}_{#1}}\xspace}

\newcommand{\csdict}[1][]{\ensuremath{\bm\Psi_{#1}}\xspace}
\newcommand{\csmeasmtx}[1][]{\ensuremath{\bm\Phi_{#1}}\xspace}
\newcommand{\cssysmtx}[1][]{\ensuremath{\mathbf A_{#1}}\xspace}

\newcommand{\noisesymb}{n}
\newcommand{\noise}[1][]{\ensuremath{\mathbf \noisesymb_{#1}}\xspace}
\newcommand{\qnoisesymb}{q}
\newcommand{\qnoise}[1][]{\ensuremath{\mathbf \qnoisesymb_{#1}}\xspace}
\newcommand{\rnoisesymb}{r}
\newcommand{\rnoise}[1][]{\ensuremath{\mathbf \rnoisesymb_{#1}}\xspace}

\DeclareMathOperator{\E}{\mathbb E}
\makeatletter
\newcommand{\expectation}[1]{%
  \@ifnextchar[{\expectation@i{#1}}{\expectation@j{#1}}%]
}
\def\expectation@i#1[#2]{%
  \ensuremath{\E_{#2}\left[#1\right]}\xspace%
}
\def\expectation@j#1{%
  \ensuremath{\E\left[#1\right]}\xspace%
}
\makeatother
\newcommand{\transpose}{\ensuremath{^\mathrm{T}}}